\documentclass[letterpaper,english,aps,prb,twocolumn,floatfix, showpacs, amsfonts, amssymb]{revtex4}
\usepackage{amsmath}
\usepackage{graphicx}
\usepackage{amssymb}

\makeatletter

\usepackage{amscd}
\usepackage{bm}

\usepackage{babel}
\makeatother
\begin{document}

\title{Magnetically-controlled impurities in quantum wires with strong Rashba
coupling}

\author{R.~G.~Pereira}

\affiliation{Instituto de F\'{\i}sica Gleb Wataghin, Unicamp, Caixa Postal 6165,
13083-970 Campinas, SP, Brazil}

\author{E.~Miranda}

\affiliation{Instituto de F\'{\i}sica Gleb Wataghin, Unicamp, Caixa Postal 6165,
13083-970 Campinas, SP, Brazil}

\date{\today}

\begin{abstract}
We investigate the effect of strong spin-orbit interaction on the
electronic transport through non-magnetic impurities in one-dimensional
systems. When a perpendicular magnetic field is applied, the electron
spin polarization becomes momentum-dependent and spin-flip scattering
appears, to first order in the applied field, in addition to the usual
potential scattering. We analyze a situation in which, by tuning the
Fermi level and the Rashba coupling, the magnetic field can suppress
the potential scattering. This mechanism should give rise to a significant magnetoresistance in the limit of large barriers.
\end{abstract}


\maketitle
The spin-orbit interaction in low-dimensional systems is a powerful
tool in the development of spintronics, the electronics that intends
to explore the electron spin to store and transport information.\cite{spintronicsreview}
In semiconductor heterostructures, the spin-orbit interaction arises
intrinsically from the asymmetry of the potential that confines the
two-dimensional electron gas (2DEG). In this context, it is usually
referred to as Rashba coupling.\cite{rashba} In addition to this,
there is also the Dresselhaus spin-orbit coupling, which is a result
of the lack of inversion-symmetry in the bulk.\cite{dresselhausspinorbit}
Both terms introduce an effective magnetic field that depends on the
in-plane momentum of the electron. The spin precession associated
with the Rashba coupling led Datta and Das to propose a spin field-effect
transistor (spin-FET).\cite{dattadas} The interest in this device
is furthered a great deal by the demonstrated possibility of tuning
the spin-orbit coupling by means of applied gate voltages.\cite{grundler}
However, the presence of impurities poses an obstacle to the spin-FET
because they scatter electrons between states with different momentum
and, consequently, randomize the spin direction. This is known as
the Elliot-Yafet mechanism of spin relaxation.\cite{elliott,yafet}
More recently, it has been pointed out that a non-ballistic spin-FET,
which would be robust against impurity scattering, can be realized
by matching the Rashba and Dresselhaus parameters.\cite{schliemannetal} 

Quantum wires are created when the propagation in the 2DEG is further
confined in one of its directions. If the wire width is comparable
to the Fermi wavelength, the transverse sub-bands of the quasi-one-dimensional
system are quantized. In the strict one-dimensional (1D) limit of
vanishing width, one can neglect the mixing between the sub-bands,
which is proportional to the transverse momentum, and recover a well-defined
spin-polarization axis.\cite{governaletzulicke} The dependence of
the electron spinors on the momentum is restored and conveniently
controlled by applying a magnetic field perpendicular to the spin-polarization
axis.\cite{pershinetal} Therefore, although impurity scattering is
enhanced in 1D systems due to interaction effects,\cite{kanefisherprl,kanefisherprb,matveevetal}
a nonmagnetic impurity is able to cause spin relaxation only in the
presence of an external magnetic field.

An effective theory of 1D conductors which includes spin-orbit interaction
as well as Zeeman splitting has been developed.\cite{morozetal,yuetal}
In this work, we address the question of how the spin splitting characteristic
of spin-orbit interaction affects the scattering off nonmagnetic impurities
in 1D systems. For zero magnetic field, we obtain the usual potential
scattering without spin flip. The application of a perpendicular magnetic
field opens a gap in the vicinity of $k=0$, where the spin subbands
are degenerate, and introduces a spin-flip scattering term. We analyze
the effect of this term in the limit of strong Rashba coupling, i.e.,
when the band splitting is comparable to the Fermi energy. In Ref.~\onlinecite{stredaseba},
it was argued that electron scattering at a potential step in the
interface between a metallic gate and the semiconductor wire can work
as a spin filter when the Fermi level coincides with the position
of the gap. We will show that, by making the same choice for the Fermi
level, the spin-flip scattering which we consider becomes the most
relevant process at low temperatures and the scattering off the impurity
can be controlled by the magnetic field. The possibility of observing
this effect in a realistic experimental situation is also discussed.

The Hamiltonian for one electron moving on the $xy$ plane subjected
to the Rashba spin-orbit interaction is\begin{equation}
H_{0}=\frac{1}{2m}\left(p_{x}^{2}+p_{y}^{2}\right)+\frac{\alpha}{\hbar}\left(\sigma_{x}p_{y}-\sigma_{y}p_{x}\right)+V\left(x\right),\label{eq:model}\end{equation}
where $\mathbf{p}$ is the momentum operator, $\alpha$ is the Rashba
coupling constant, $\bm{\sigma}$ is the vector of Pauli matrices,
and $V\left(x\right)$ is a potential that confines the electron in
the $x$ direction. The latter is usually taken as the potential of
a harmonic oscillator $V\left(x\right)=m\omega_{0}^{2}x^{2}/2$, and
the 1D limit is achieved by taking $\hbar\omega_{0}\gg\epsilon_{F}$,
with $\epsilon_{F}$ the Fermi energy. This means that the wire width
$w$ must be small enough that $w\ll\lambda_{F}$, where $\lambda_{F}$
is the Fermi wavelength. In this limit, the transverse degrees of
freedom are frozen at low temperatures and we can simply write\begin{equation}
H_{0}\approx\frac{p^{2}}{2m}+\frac{\alpha}{\hbar}\sigma_{x}p+\frac{\hbar\omega_{0}}{2},\label{eq:H1DB=0}\end{equation}
where $p\equiv p_{y}$ is the component of the momentum in the direction
of the wire and $\hbar\omega_{0}/2$ is the zero point energy (lowest
subband) of the oscillator in the transverse direction. We also include
in the model the effect of a magnetic field applied along the $z$-direction\begin{equation}
H_{0}\approx\frac{p^{2}}{2m}+\frac{\alpha}{\hbar}\sigma_{x}p-\frac{g\mu_{B}B}{2}\,\sigma_{z}+\frac{\hbar\omega_{0}}{2}.\label{eq:H1D}\end{equation}
Here, $g$ is the effective electron g-factor, $\mu_{B}$ is the Bohr
magneton, and $B$ is the applied magnetic field. As usual, we ignore
the orbital effect of the magnetic field in the 1D limit. Actually,
what really matters for our purposes is that the external field is
perpendicular to the spin direction set by the spin-orbit coupling.
Thus, a field applied along the $y$-direction (which causes no orbital
effect) would serve as well.

\begin{figure}
\begin{center}\includegraphics[%
  scale=0.3]{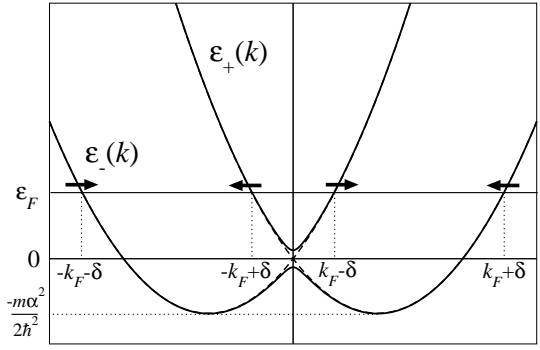}\end{center}

\caption{\label{cap:fig1}Electron dispersion in the presence of spin-orbit
interaction, for $B=0$ (dashed line) and $B\neq0$ (solid line).
If $\alpha k_{F}\gg g\mu_{B}B$, the electron states at the Fermi
surface have spins approximately parallel to $+x$ ($\rightarrow$)
or $-x$ ($\leftarrow$). }
\end{figure}

The Hamiltonian $H_{0}$ in Eq.~(\ref{eq:H1D}) is promptly diagonalized
in momentum space. For $B=0$, the eigenfunctions are plane waves
with eigenvalues (omitting the zero point energy) $\epsilon_{\sigma}=\hbar^{2}k^{2}/2m+\sigma\alpha k$,
where $\sigma=+$ ($\sigma=-$) refers to spins parallel to $+x$
($-x$). As a result, the spin bands are {}``horizontally'' split
even at $B=0$. The energy minima are shifted to $k=\pm m\alpha/\hbar^{2}\equiv\pm\delta$
and the bands are degenerate at $k=0$ (Fig.~\ref{cap:fig1}).

However, for $B\neq0$, the eigenfunctions take the form $\Psi_{\lambda}\left(x\right)=e^{ikx}\chi_{\lambda}\left(k\right)$,
$\lambda=\pm$, where the momentum-dependent spinors are written in
the basis of eigenstates of $\sigma_{z}$ as\begin{equation}
\chi_{+}\left(k\right)=\left(\begin{array}{c}
\sin\frac{\theta\left(k\right)}{2}\\
\cos\frac{\theta\left(k\right)}{2}\end{array}\right),\,\,\,\chi_{-}\left(p\right)=\left(\begin{array}{c}
\cos\frac{\theta\left(k\right)}{2}\\
-\sin\frac{\theta\left(k\right)}{2}\end{array}\right).\label{eq:spinors}\end{equation}
Here, we have introduced the polar angle $\theta\left(k\right)$ that
specifies the polarization direction in the $xz$ plane, given by\begin{equation}
\theta\left(k\right)=\tan^{-1}\frac{2\alpha k}{g\mu_{B}B}.\end{equation}
The eigenvalues of $H_{0}$ are \begin{equation}
\epsilon_{\pm}\left(k\right)=\frac{\hbar^{2}k^{2}}{2m}\pm\sqrt{\left(\alpha k\right)^{2}+\left(\frac{g\mu_{B}B}{2}\right)^{2}}.\label{eq:dispersion}\end{equation}
The dispersion for $B\neq0$ is shown in Fig.~\ref{cap:fig1}. We
note that the mixing between $\sigma=\pm$ eigenstates opens a gap
$\Delta=g\mu_{B}B$ at $k=0$. Furthermore, we can no longer associate
a fixed spin direction with the bands $\lambda=\pm$. Consider, for
example, the limit of strong Rashba coupling $\alpha k_{F}\gg g\mu_{B}B$,
where $k_{F}=n\pi/2$, with $n$ the average electron density, stands
for the Fermi momentum when $\alpha=0$. For $\epsilon_{F}>0$ (as
in Fig.~\ref{cap:fig1}), the electron states with $\lambda=-$ on
the Fermi surface have spin approximately parallel to $+x$ at $k=-k_{F}-\delta$
and $-x$ at $k=+k_{F}+\delta$. The opposite happens for $\lambda=+$.
As one moves from one Fermi point to the other along each band, the
spin polarization changes continuously from $\pm x$ to $\mp x$,
passing through $\pm z$ in the vicinity of the gap, where $\left|k\right|\ll g\mu_{B}B/\alpha$. 

We consider now the scattering of electrons described by the spin
states in Eq.~(\ref{eq:spinors}) off nonmagnetic impurities. The
free Hamiltonian of the electron gas can be written in second-quantized
form\begin{equation}
H_{0}=\sum_{k,\lambda}\epsilon_{\lambda}\left(k\right)\, c_{k\lambda}^{\dagger}c_{k\lambda}^{\phantom{\dagger}},\end{equation}
where $\epsilon_{\lambda}\left(k\right)$ are the dispersions in Eq.~(\ref{eq:dispersion})
and $c_{k\lambda}$ destroys an electron in the eigenstate $\left|k\lambda\right\rangle $
with momentum $k$ in the band $\lambda$. The nonmagnetic impurity
corresponds to the perturbation\begin{equation}
V=\sum_{kp\lambda\mu}\left\langle k\lambda\left|V\left(x\right)\right|p\mu\right\rangle c_{k\lambda}^{\dagger}c_{p\mu},\label{eq:impurity}\end{equation}
where $V\left(x\right)$ is a localized potential. Using the spinors
in Eq.~(\ref{eq:spinors}), we can calculate the matrix elements
in Eq.~(\ref{eq:impurity}) to find\begin{eqnarray}
V & = & \sum_{kp\lambda}\frac{V_{k-p}}{N}\cos\left[\frac{\theta\left(k\right)-\theta\left(p\right)}{2}\right]\, c_{k\lambda}^{\dagger}c_{p\lambda}\nonumber \\
 & - & \sum_{kp\lambda}\lambda\frac{V_{k-p}}{N}\sin\left[\frac{\theta\left(k\right)-\theta\left(p\right)}{2}\right]\, c_{k\lambda}^{\dagger}c_{p,-\lambda}.\label{eq:doistermos}\end{eqnarray}
where $V_{k}$ is the the Fourier transform of $V\left(x\right)$
and $N$ is the number of sites in the wire. The first term in Eq.~(\ref{eq:doistermos})
describes scattering between states in the same band whereas the second
one describes scattering between states in different bands. That these
two terms exist in general is a feature of the spin-orbit interaction,
which makes $\theta\left(k\right)\neq\theta\left(p\right)$ if $k\neq p$.
Since a given band does not have a fixed spin direction, spin-flip
scattering terms can emerge from both terms. In order to make these
terms explicit, we focus on the limit of $g\mu_{B}B\ll\alpha k$ for
$k$ around the Fermi surface in Fig.~\ref{cap:fig1} and expand
$\theta\left(k\right)\approx\textrm{sgn}\left(k\right)\pi/2-g\mu_{B}B/2\alpha k$.
To zeroth order in the magnetic field, we find\begin{eqnarray}
V^{\left(0\right)} & = & \sum_{kp\lambda}\frac{V_{k-p}}{N}\,\left[\frac{1+\textrm{sgn}\left(kp\right)}{2}\, c_{k\lambda}^{\dagger}c_{p\lambda}\right.\nonumber \\
 & - & \left.\lambda\,\frac{\textrm{sgn}\left(k\right)-\textrm{sgn}\left(p\right)}{2}\, c_{k\lambda}^{\dagger}c_{p,-\lambda}\right].\label{eq:Vordemzero}\end{eqnarray}
It is easy to verify that the two terms in Eq.~(\ref{eq:Vordemzero})
correspond to the usual potential scattering in the limit $B\rightarrow0$.
It is possible to scatter between states in the same band and with
momenta of the same sign or states in different bands and with momenta
of opposite signs. According to Fig.~\ref{cap:fig1}, only the latter
case can occur if $\epsilon_{F}>0$ and it consists of a non spin-flip
process that connects $k_{F}\pm\delta$ with $-k_{F}\pm\delta$. True
spin-flip processes appear at first order in $B$\begin{eqnarray}
V^{\left(1\right)} & = & \frac{g\mu_{B}B}{4\alpha}\sum_{kp\lambda}\frac{V_{k-p}}{N}\,\left(\frac{1}{k}-\frac{1}{p}\right)\nonumber \\
 & \times & \left[\frac{\textrm{sgn}\left(k\right)-\textrm{sgn}\left(p\right)}{2}\, c_{k\lambda}^{\dagger}c_{p\lambda}\right.\nonumber \\
 & - & \left.\lambda\,\frac{1+\textrm{sgn}\left(kp\right)}{2}\, c_{k\lambda}^{\dagger}c_{p,-\lambda}\right].\label{eq:vfip}\end{eqnarray}
The terms in $V^{\left(1\right)}$ connect $k_{F}\pm\delta$ with
$-\left(k_{F}\pm\delta\right)$, which are states with spins polarized
in opposite directions ($\pm x$). The momentum transfer in the scattering
is $\Delta k=2\left(k_{F}\pm\delta\right)$.

In the limit of strong Rashba coupling and low magnetic field, the
potential scattering off the impurity overcomes the spin-flip scattering
because the amplitude of the latter is proportional to $g\mu_{B}B/\alpha k_{F}\ll1$.
Hence, we do not expect a large variation of the resistance of the
wire with the magnetic field. However, the band structure in Fig.~\ref{cap:fig1}
suggests a special case in which we can suppress the potential scattering.
As first proposed in Ref.~\onlinecite{stredaseba}, it suffices to
lower the Fermi energy until $\epsilon_{F}=0$, so that the Fermi
level lies exactly inside the gap. In this case, the states to which
electrons would be scattered without spin flip become forbidden and
this process will depend on thermal activation to states above the
gap. Consequently, the spin-flip scattering will dominate at temperatures
much lower than the gap, which is controlled by the external magnetic
field. The condition $\epsilon_{F}=0$ relies on a fine tuning between
the Rashba coupling and the electronic density, and can be stated
as $\alpha=\hbar v_{F}$, where $v_{F}=\hbar k_{F}/m$ is the Fermi
velocity.

\begin{figure}
\begin{center}\includegraphics[%
  scale=0.3]{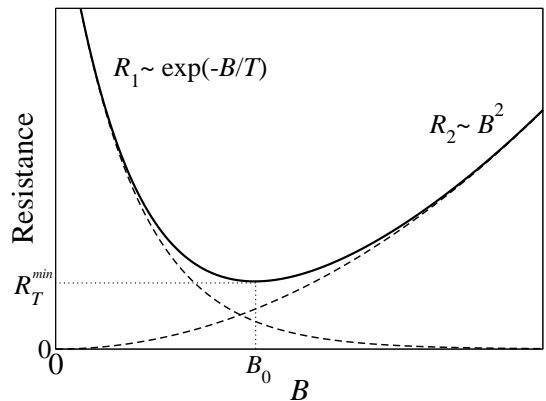}\end{center}

\caption{\label{cap:fig2}Resistance of the quantum wire with Rashba coupling
constant $\alpha=\hbar v_{F}$, for a fixed temperature and as a function
of the magnetic field $B$.}
\end{figure}

We calculate the resistance for this situation perturbatively by using
the memory function formalism.\cite{gotzewolfle} First, we note that
even the {}``usual'' scattering (in the sense that it reduces to
the potential scattering in the limit $B\rightarrow0$) from $k=\pm\left(k_{F}+\delta\right)\equiv\pm2k_{F}$
to $k\approx0$ involves a rotation of $\pm\pi/2$ in the spin direction.
For $k_{B}T\ll\Delta\ll \alpha k_F$, the resistance due to this scattering is\begin{equation}
R_{1}=\frac{h}{e^{2}}\left(\frac{ma\left|V_{2k_{F}}\right|}{\pi\hbar^{2}n}\right)^{2}\sqrt{\frac{\pi g\mu_{B}B}{k_{B}T}}\exp\left(-\frac{g\mu_{B}B}{2k_{B}T}\right).\end{equation}
where $a$ is the lattice parameter. Likewise, the resistance due
to the spin-flip scattering ($\Delta k=4k_{F}$) reads\begin{equation}
R_{2}=\frac{h}{e^{2}}\left(\frac{ma\left|V_{4k_{F}}\right|}{\pi\hbar^{2}n}\right)^{2}\left(\frac{g\mu_{B}B}{4\alpha k_{F}}\right)^{2}.\end{equation}
We see that $R_{1}$ decreases and $R_{2}$ increases as the magnetic
field is increased. In principle, we should be able to observe a crossover
to a regime in which the total resistance is due mostly to the spin-flip
scattering (Fig.~\ref{cap:fig2}). Moreover, there must be a total
resistance minimum for a given value of $B=B_{0}$. Let us estimate
this minimum resistance $R_{T}^{min}$. By putting $V_{2k_{F}}\approx V_{4k_{F}}$,
$T=1\,\textrm{K}$, $k_{F}=2\times10^{6}\,\textrm{cm}^{-1}$, $m=0.036\, m_{0}$
(for InAs quantum wells\cite{grundler}), where $m_{0}$ is free electron
mass, and $g=13$ (Ref.~\onlinecite{brosigetal}), we get $B_{0}\approx3.2\,\textrm{T}$
and \[
\frac{R_{T}^{min}}{R_{T}^{0}}\approx3\times10^{-5},\]
where $R_{T}^{0}=2\left(h/e^{2}\right)\left(ma\left|V_{2k_{F}}\right|/\pi\hbar^{2}n\right)^{2}$
is the resistance associated with the impurity scattering at $B=0$.
In practice, this mechanism removes the impurity scattering at low
temperatures. 

We define the magnetoresistance as the relative difference between the resistance at $B=B_0$ and the resistance at zero magnetic field\begin{eqnarray}
\frac{\Delta R}{R} & = & \frac{\left(G_{0}^{-1}+R_{T}^{min}\right)-\left(\frac{1}{2}G_{0}^{-1}+R_{T}^{0}\right)}{\left(G_{0}^{-1}+R_{T}^{min}\right)}\nonumber \\
 & \approx & \frac{1}{2}-\frac{1}{2}\left(\frac{ma\left|V_{2k_{F}}\right|}{\hbar^{2}k_{F}}\right)^{2},\label{eq:MR}\end{eqnarray}
where $G_{0}=e^{2}/h$ is the conductance quantum. In the limit of
weak barriers ($\left|V_{2k_{F}}\right|\rightarrow0$), the magnetoresistance
is positive because the main effect of the magnetic field is to open
the gap at $k=0$ and reduce the number of electron channels at the
Fermi level to the equivalent of a single band ($\lambda=-$). As
a result, the ballistic conductance decreases, according to the Landauer
formula,\cite{dattabook} from $2G_{0}$ to $G_{0}$ (Ref.~\onlinecite{pershinetal}).
However, in the limit of large barriers ($R_{T}^{0}\gg G_{0}^{-1}$),
the magnetoresistance is negative because, upon effectively removing
the impurity with the magnetic field, we increase the conductance
from almost zero to $G_{0}$. Actually, this is the most relevant
case in the 1D limit, since repulsive interactions renormalize upwards
the scattering off an impurity in a Luttinger liquid.\cite{kanefisherprl,kanefisherprb,matveevetal}
Therefore, the scattering amplitude $\left|V_{2k_{F}}\right|$ in
Eq.~(\ref{eq:MR}) should be replaced by the effective amplitude
$\left|V_{2k_{F}}\right|\left(T/T_{F}\right)^{K-1}$, where $K<1$
for repulsive interactions, leading to singular scattering at $T=0$.
For this reason, a large negative magnetoresistance should be observed.
It must also be remarked that at $T=0$ the spin-flip scattering is
the only source of electron scattering. Since this process converts
right-moving {}``spin-down'' ($\leftarrow$) electrons into left-moving
{}``spin-up'' ($\rightarrow$) electrons, and vice-versa, it conserves
the persistent spin current $\left\langle J_{s}\right\rangle =\left\langle J_{\rightarrow}-J_{\leftarrow}\right\rangle \neq0$
in a wire with spin-orbit interaction. On the other hand, it does not conserve spin density because the magnetic field restores the Elliot-Yafet mechanism of spin-relaxation.

Finally, we consider the possibility of observing this effect with
the available experimental conditions. For a density as low as $k_{F}=2\times10^{6}\,\textrm{cm}^{-1}$
and for $m=0.036\, m_{0}$, we need $\alpha\approx4\times10^{-10}\,\textrm{eVm}$,
which is one order of magnitude above the reported values of $\alpha$
(Ref.~\onlinecite{grundler}). However, the condition $\alpha=\hbar v_{F}$
could be achieved in principle by either further reducing the electron
density or pursuing higher values of $\alpha$, by means of applied
electric fields or more asymmetric heterostructures. Besides, it has
been predicted that interaction effects should enhance $\alpha$ at
low densities.\cite{hausler} The fine tuning between the Rashba splitting
and the Fermi level (with an error smaller than the gap $\Delta$)
could also be guaranteed, since it is possible to control these two
properties independently.\cite{grundler}

We would like to end by mentioning an interesting recent analysis of
the magnetoresistance of quantum wires in the presence of a domain
wall and spin-orbit interaction.\cite{dugaevetal} Unlike us, however,
that work focuses on the limit where the effective magnetic field
``seen'' by the conduction electrons (produced by the polarized spin
background) is much greater than the spin-orbit scale:
$g\mu_{B}B\gg\alpha k_{F}$.  Furthermore, the tuning of the Fermi
level to the gap is not considered. The role of the chemical potential
when there is a gap due to a perpendicular magnetic field has also
appeared in a related work.\cite{nesteroffetal04} It focuses on a
combination of spin-orbit interaction and the time-dependent magnetic
field provided by the coupling to a non-equilibrium nuclear spin
polarization, which is predicted to induce fluctuations in the charge
conductance.\cite{nesteroffetal04}

In conclusion, we have shown that the scattering off nonmagnetic impurities
is strongly affected by a perpendicular magnetic field when the Fermi
level crosses the degeneracy point of 1D bands split by spin-orbit
interaction. The opening of a gap proportional to the Zeeman energy
suppresses the potential scattering at low temperatures. This mechanism
does not depend on the scattering amplitude. Therefore, one can effectively
remove all the impurities in the wire with the magnetic field and
observe a negative magnetoresistance.

The authors acknowledge financial support by Fapesp through grants
01/12160-5 (R. G. P.) and 01/00719-8 (E. M.), and by CNPq through
grant 301222/97-5 (E. M.).

\end{document}